\documentclass{article}
\usepackage{graphicx}
\usepackage{amsmath}
\title{On classification of higher-order integrable nonlinear partial differential equations}
\date{}
\author{Il'in I.A.,Noshchenko D.S.,Perezhogin A.S.
\\ \ \\ \small
IKIR FEB RAS, 684034 Kamchatka region, Elizovskiy district, Paratunka, Mirnaya str., 7.
\\ \small
KamGU Vitus Bering, 683032, Kamchatka region, Elizovskiy district, \\
\small Petropavlovsk-Kamchatsky, Pogranichnaya str., 4.}
\begin{document}

\maketitle


\begin{abstract}

In the paper we investigate existence of soliton solutions for higher-order partial differential
equations with polynomial nonlinearities.
Using the $\tau$-function method we obtain classification for high-order integrable
systems.

\end{abstract}

%
%

\section{Introduction}

In nonlinear physics solitons are defined as solitary waves which interact in elastic way, like particles
.


Classical example -- the Kortevieg de Vries (KdV) equation
\begin{equation}\label{0}
u_t+6uu_x+u_{xxx}=0
\end{equation}
admits $n$-soliton solution that can be expressed in terms of Hirota's $\tau$-function.

Consider some $n$-th order partial differential equation with polynomial nonlinearities in
1+1 variables $E(u,u_t,u_xu,u_{xx}u_x,{u_x}^2,\dots)$.
Let $u(x,t)$ its meromorphic solution (only finite number of negative powers contained in its Laurent expansion)
\begin{equation}{\label{sum}}
u(x,t)=\sum_{k=-p}^{\infty}c_k\phi(x,t)^k
\end{equation}
Integer $p$ is called a singular order of equation \cite{Goriely, Conte}. It can be obtained
by substitution $u(\xi)=\frac{1}{\xi^p}$, where $\partial_\xi\sim\partial_x$.
For the KdV equation \ref{0} we have
\begin{equation}
{\frac {d^{3}}{d{\xi}^{3}}}u \left( \xi \right) +6\, \left( {\frac {d}
{d\xi}}u \left( \xi \right)  \right) u \left( \xi \right) = (p^2+3p+2)\xi^{-p-3}+6\xi^{-2p-1}
\end{equation}
To equal poles we take $p+3=2p+1$, so $p=2$.

We denote $n$-th order partial diffenrential equation
of singular order $p$ as $E_p^n(u)$.
Hirota's $\tau$-function \cite{Hirota, Conte} for $E_p^n(u)$ equation is defined as
\begin{equation}\label{hsubs}
u(x,t)=K\frac{\partial^p}{\partial{}x^p}\log\tau(x,t)
\end{equation}
For example, for the KdV equation one-soliton solution (solitary wave) can be expressed as
\begin{equation}
u(x,t)=\frac{\partial^2}{\partial{}x^2}\log(1+\exp(px-qt))
\end{equation}
and two-soliton solution
\begin{equation}
u(x,t)=\frac{\partial^2}{\partial{}x^2}\ln(1+\exp(p_1x-q_1t)+\exp(p_2x-q_2t)+\alpha_{12}\exp((p_1+p_2)x-(q_1+q_2)t))
\end{equation}
where $\alpha_{12}=\left(\frac{p_1-p_2}{p_1+p_2}\right)^2$, $p_i,q_i$ are constants.
In general, for $n$-soliton solution we have
\begin{equation}
u(x,t)=\frac{\partial^2}{\partial{}x^2}\log\left(1+\sum_{i=1}^{n}f_i(x,t)\right)
\end{equation}
where $f_i(x,t)$ is a polynomial in $e^{p_jx-q_jt}$ for $j=1..n$. We say that the equation is $\emph{partially integrable}$ if it has two-soliton solution in Hirota's form.
A remarkable fact is that some equations with soliton-type solutions (we call them soliton equations)
form a hierarchies (Lax, SK, KK equations), i.e. they admits the same type of solitons. For example,
famous KdV equation is the first element in Lax hierarchy.

Our primary goal is to investigate existence of high order $E_2^n$ equations with two-soliton solutions \cite{Ablowitz, Hirota, Nuseir}. All of these equations have terms with $\partial_x$ of singular order $p=2$ and only one term with $\partial_t$
\\
\
\\
\begin{tabular}{l l}
$E^3_2(u)= $&$u_t+a_1uu_x+u_{xxx}$\\
\\
$E^5_2(u)=$&$u_t+a_1uu_{xxx}+a_2u^2u_{x}+a_3u_xu_{xx}+u_{xxxxx}$\\
\\
$E^7_2(u)=$&$u_t+a_1u^3_x+a_2uu_xu_{xx}+a_3uu_{xxxxx}+a_4u^2u_{xxx}+a_5u^3u_x+a_6u_xu_{xxxx}+$
\\
\\
&$+a_7u_{xx}u_{xxx}+u_{xxxxxxx}$\\
$\vdots$
\end{tabular}
\\
\
\\
and so on, for arbitary large $n$.

\section{Soliton solutions for $E^n_2$}

With $\tau$-function expansion
\begin{equation}\label{heps}
\tau=1+\varepsilon{}f_1+\varepsilon^2{}f_2+\varepsilon^3{}f_3+\dots
\end{equation}
where $\varepsilon$ is an arbitrary complex parameter,
we obtain substitution for two-solition solution \cite{Hirota, Nuseir, Kudryashov_1}:
\begin{eqnarray}
\label{fsys1}
f_1(x,t)=\exp(p_1x-q_1t)+\exp(p_2x-q_2t)\\
\label{fsys2}
f_2(x,t)=\alpha_{12}\exp((p_1+p_2)x-(q_1+q_2)t)\\
\label{fsys3}
f_{i}(x,t)=0, i\geq{}3
\end{eqnarray}
$\alpha_{12}=\frac{R_n(p_1,q_i,p_2,q_j)}{S_n(p_1,q_i,p_2,q_j)}$ is a rational function of complex parameters $p_1,q_i$.
After substituting \ref{fsys1}-\ref{fsys3} at $E_2^n$
we equal coefficients at $\varepsilon$. The resulting algebraic system depends on $a_i,\alpha_{12}$ coefficients.
Evaluation of linear part $u_{nx}+u_t$ leads to $q_i=p^n_i$, which is called a dispersion relation \cite{Nuseir}.

Now we are going to obtain explicit form for denominator $S_n$ of $\alpha_{12}$. With
\begin{equation}\label{Dseq}
\frac{\partial^n}{\partial{}x^n}+\mathcal{I}\frac{\partial}{\partial{}x}+\frac{\partial}{\partial{}t}
, \quad n=3,5,7,\dots
\end{equation}
where $\mathcal{I}u=u$ and $u(x,t)$ is 2-soliton substitution \ref{heps} with 9-11, we equal coefficients at $\varepsilon$.
Evaluating $\varepsilon^2$ coefficient from the expansion above, we get polynoms $S_n$ of the form
\\
\
\\
\begin{tabular}{ll}
$S_3=$&$3(p_1+p_2)^2$ \\
\\
$S_5=$&$5(p_1+p_2)^2(p_1^2+p_1p_2+p_2^2)$ \\
\\
$S_7=$&$7(p_1+p_2)^2(p_1^2+p_1p_2+p_2^2)^2$ \\
\\
$S_9=$&$3\left( p_{{i}}+p_{
{j}} \right) ^{2}\left( 3\,{p_{{i}}}^{6}+9\,{p_{{i}}}^{5}p_{{j}}+19\,{p_{{i}}}^{4}{p_{
{j}}}^{2}+23\,{p_{{i}}}^{3}{p_{{j}}}^{3}+19\,{p_{{i}}}^{2}{p_{{j}}}^{4
}+9\,p_{{i}}{p_{{j}}}^{5}+3\,{p_{{j}}}^{6} \right)$\\
\\
$S_{11}=$&$11  \left( p_{{i}}+p_{{j}} \right) ^{2}\left( {p_{{i}}}^{2}+p_{{i}}p_{{j}}+{p_{{j}}}^{2} \right)  \left( {p_
{{j}}}^{6}+3\,p_{{i}}{p_{{j}}}^{5}+7\,{p_{{i}}}^{2}{p_{{j}}}^{4}+9\,{p
_{{i}}}^{3}{p_{{j}}}^{3}+7\,{p_{{i}}}^{4}{p_{{j}}}^{2}+3\,{p_{{i}}}^{5
}p_{{j}}+{p_{{i}}}^{6} \right)
$\\
\\
$S_{13}=$&$13  \left( p_{{i}}+p_{{j}}
 \right) ^{2} \left( {p_{{i}}}^{2}+p_{{i}}p_{{j}}+{p_{{j}}}^{2}
 \right) ^{2}\, \left( {p_{{j}}}^{6}+3\,p_{{i}}{p_{{j}}}^{5}+8\,{p_{{i}}}^{2}{p_{
{j}}}^{4}+11\,{p_{{i}}}^{3}{p_{{j}}}^{3}+8\,{p_{{i}}}^{4}{p_{{j}}}^{2}
+3\,{p_{{i}}}^{5}p_{{j}}+{p_{{i}}}^{6} \right)
$\\
\\
$S_{15}=$&$ \left( p_{{i}}+p
_{{j}} \right) ^{2}( 15\,{p_{{i}}}^{12}+15\,{p_{{j}}}^{12}+90\,p_{{j}}{p_{{i}}}^{11
}+90\,{p_{{j}}}^{11}p_{{i}}+365\,{p_{{j}}}^{10}{p_{{i}}}^{2}+1000\,{p_
{{j}}}^{9}{p_{{i}}}^{3}+2003\,{p_{{j}}}^{8}{p_{{i}}}^{4}+$\\
\\
 &$
+3002\,{p_{{j}
}}^{7}{p_{{i}}}^{5}+3433\,{p_{{j}}}^{6}{p_{{i}}}^{6}+3002\,{p_{{j}}}^{
5}{p_{{i}}}^{7}+2003\,{p_{{j}}}^{4}{p_{{i}}}^{8}+1000\,{p_{{j}}}^{3}{p
_{{i}}}^{9}+365\,{p_{{j}}}^{2}{p_{{i}}}^{10} )  
$\\
\\
$\vdots$ \\
\end{tabular}
\\
\
\\
These polynomials can be derived by horizontal summation of neighbour elements in a binomial triangle, as shown below.
\\
\
\\
\begin{tabular}{c}
                                    \textbf{1}, \textbf{1} \\
                                 \textbf{1}, 3, 3 \textbf{1}\\
                             \textbf{1}, 5, 10, 10, 5, \textbf{1}\\
                         \textbf{1}, 7, 21, 35, 35, 21, 7, \textbf{1}\\
                    \textbf{1}, 9, 36, 84, 126, 126, 84, 36, 9, \textbf{1}\\
             \textbf{1}, 11, 55, 165, 330, 462, 462, 330, 165, 55, 11, \textbf{1}\\
      \textbf{1}, 13, 78, 286, 715, 1287, 1716, 1716, 1287, 715, 286, 78, 13, \textbf{1}\\
      \vdots \\
      $\Downarrow$\\
                                   3, 6, 3\\
                              5, 15, 20, 15, 5\\
                          7, 28, 56, 70, 56, 28, 7\\
                    9, 45, 120, 210, 252, 210, 120, 45, 9\\
              11, 66, 220, 495, 792, 924, 792, 495, 220, 66, 11\\
     13, 91, 364, 1001, 2002, 3003, 3432, 3003, 2002, 1001, 364, 91, 13\\
  \vdots
\end{tabular}  
\\
\
\\
For every integer $n$ roots of $S_n$
lie on curve defined this way: $Re(z)\in\left(-1,-\frac{1}{2}\right)$ implies unit circle
centered at $(0,0)$, $Re(z)\in\left(-\frac{1}{2},0\right)$ implies unit circle
centered at $(-1,0)$, $Re(z)=-\frac{1}{2}$ implies line $Re(z)=\frac{1}{2}$ minus
segment bounded by the first two circles. At last, $Re(z)=-1,0$ are limit
points when $n\to\infty$.
\begin{figure}[h!]
\centering
\includegraphics[scale=0.3]{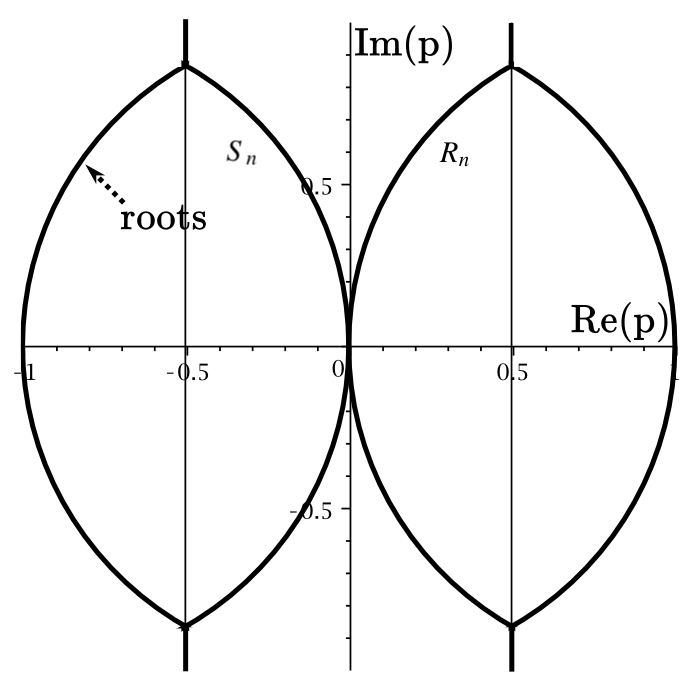}
\caption{Root geometry of $S_n$ and $R_n$ polynomials on complex plane}
\end{figure}

Now we should define numerator $R_n$ of $\alpha_{12}$. Let
\begin{equation}
S_n=(p_1+p_2)^2(p_1^2+p_1p_2+p_2^2)^{k}T_n(p_1,p_2), k=0,1,2
\end{equation}
Direct computation gives two variants
\begin{equation}
R_n(p_1,p_2)=S_n(-p_1,p_2) \label{tail}\\
\end{equation}
and
\begin{equation}\label{notail}
R_n(p_1,p_2)=\frac{S_n(-p_1,p_2)(p_1^2+p_1p_2+p_2^2)^lT_n(p_1,p_2)}{(p_1^2-p_1p_2+p_2^2)^lT_n(-p_1,p_2)}, l\leq{}k
\end{equation}
These selections keep symmetry property of $\alpha_{12}$, so its zeros are reflected poles by $Im(z)$ axis.
This principle allows equation $(E^n_p)_{\alpha_{12}}(u)$ with two-soliton solution
in form (\ref{heps}) to be determined explicitly. It is uniqely defined by $\alpha_{12}$. In case of (\ref{tail}) this equation 
has no hierarchy. It can be constructed for any $n\geq{}9$.
Another case (\ref{notail}) leads to widely known completely integrable Lax and SK hierarchies ($l=2,1$ respectively)
\cite{Goriely}. When $l=0$, we get new hierarchy ($n=7,13,19\dots$).
For $n=7$, for example, this is equation (\ref{new}). (\ref{new_flux}) represents its flux $J_1$, associated with conserving
density $\rho_1=u$ \cite{Nuseir, Gotkas}.
We also got that new and tail equations do not have a three-soliton solution.

Summary result about classification of equations and its properties are represented at table 1.

\begin{tabular}{l|ll}
Lax, $n=3,5,7,\dots$ & $\alpha_{12}=$ &$\left(\frac{p_1-p_2}{p_1+p_2}\right)^2$\\
\\
SK, $n=5,7,11,\dots$ & $\alpha_{12}=$ &$\left(\frac{p_1-p_2}{p_1+p_2}\right)^2\left(\frac{p_1^2-p_1p_2+p_2^2}{p_1^2+p_1p_2+p_2^2}\right)$\\
\\
new, $n=7,13,19,\dots$ & $\alpha_{12}=$ &$\left(\frac{p_1-p_2}{p_1+p_2}\right)^2\left(\frac{p_1^2-p_1p_2+p_2^2}{p_1^2+p_1p_2+p_2^2}\right)^2$\\
\\
tail, $n=9$ & $\alpha_{12}=$ &$ \left(\frac{p_1-p_2}{p_1+p_2}\right)^2 \left(\frac{ 3\,{p_1}^{6}-9\,{p_1}^{5}{
\it p_2}+19\,{{\it p_2}}^{2}{p_1}^{4}-23\,{p_1}^{3}{{\it p_2}}^{3}+19\,{{\it 
p_2}}^{4}{p_1}^{2}-9\,p_1{{\it p_2}}^{5}+3\,{{\it p_2}}^{6}  }{
 3\,{p_1}^{6}+9\,{p_1}^{5}{\it p_2}+19\,{{\it p_2}}^{2}{p_1}^{4}+23\,{p_1
}^{3}{{\it p_2}}^{3}+19\,{{\it p_2}}^{4}{p_1}^{2}+9\,p_1{{\it p_2}}^{5}+3\,{{
\it p_2}}^{6} }\right)$\\
\end{tabular}
\\
\
\\
\begin{table}[h!]\label{class}
\begin{tabular}{|c|c|c|c|c|}
n & Lax & SK & new & tail\\
3 & $*$ & $-$ & $-$ & $-$ \\
5 & $*$ & $*$ & $-$ & $-$ \\
7 & $*$ & $*$ & $*$ & $-$ \\
9 & $*$ & $-$ & $-$ & $*$ \\
11 & $*$ & $*$ & $-$ & $*$ \\
13 & $*$ & $*$ & $*$ & $*$ \\
15 & $*$ & $-$ & $-$ & $*$ \\
17 & $*$ & $*$ & $-$ & $*$ \\
19 & $*$ & $*$ & $*$ & $*$ \\
21 & $*$ & $-$ & $-$ & $*$ \\
\end{tabular}
\quad \quad \quad \quad
\begin{tabular}{r|c|c|c|c|}
 & Lax & SK & new & tail\\
2-soliton sol. & $*$ & $*$ & $*$ & $*$ \\
3-soliton sol. & $*$ & $*$ & $-$ & $-$ \\
hierarchy & $*$ & $*$ & $*$ & $-$
\end{tabular}
\caption{$E^n_2$ classification; $*$ denotes existence of an equation at given $n$, $*$
denotes existence of given property}
\end{table}

We established the fact that some hierarchies are ``holey'': for example, SK equation of 9th order does not exist.

\begin{equation}\label{new}
\begin{aligned}
(\mathrm{new})^7_2(u)= & u_{{x,x,x,x,x,x,x}}+u_{{t}}+56\,{\frac {uu_{{x,x,x,x,x}}}{K}}+840\,{\frac {{u}^{2}u_{{x,x,x}}}{{K}
^{2}}}+3360\,{\frac {{u}^{3}u_{{x}}}{{K}^{3}}}+
\\
& +140\,{\frac {u_{{x,x,x}
}u_{{x,x}}}{K}}+56\,{\frac {u_{{x,x,x,x}}u_{{x}}}{K}}+
1680\,{\frac {uu_{{x,x}}u_{{x}}}{{K}^{2}}}
\end{aligned}
\end{equation}

\begin{equation}\label{new_flux}
J_1(u)=56\,{\frac {uu_{{x,x,x,x}}}{K}}+840\,{\frac {{u}^{4}}{{K}^{3}}}+840\,{
\frac {{u}^{2}u_{{x,x}}}{{K}^{2}}}+70\,{\frac {{u_{{x,x}}}^{2}}{K}}+u_
{{x,x,x,x,x,x}}
\end{equation}

\begin{math}
(\mathrm{tail})^9_2(u)=
420\,{\frac {u_{{x,x,x}}u_{{x,x,x,x}}}{K}}+90\,{\frac {uu_{{x,x,x,x,x,
x,x}}}{K}}+420\,{\frac {u_{{x,x,x,x,x}}u_{{x,x}}}{K}}+90\,{\frac {u_{{
x}}u_{{x,x,x,x,x,x}}}{K}}+u_{{t}}+u_{{x,x,x,x,x,x,x,x,x}}+2520\,{
\frac {{u}^{2}u_{{x,x,x,x,x}}}{{K}^{2}}}
+25200\,{\frac {{u}^{3}u_{{x,x
,x}}}{{K}^{3}}}+75600\,{\frac {{u}^{4}u_{{x}}}{{K}^{4}}}+
6300\,{\frac 
{u_{{x}}{u_{{x,x}}}^{2}}{{K}^{2}}}
+5040\,{\frac {uu_{{x}}u_{{x,x,x,x}}
}{{K}^{2}}}
+12600\,{\frac {uu_{{x,x}}u_{{x,x,x}}}{{K}^{2}}}+75600\,{
\frac {{u}^{2}u_{{x}}u_{{x,x}}}{{K}^{3}}}
\end{math}

\begin{math}
(\mathrm{tail})^{11}_2(u)=
110880\,{\frac {{u}^{3}u_{{x,x,x,x,x}}}{{K}^{3}}}+1848\,{\frac {u_{{x,
x,x,x,x}}u_{{x,x,x,x}}}{K}}+55440\,{\frac {uu_{{x,x,x,x}}u_{{x,x,x}}}{
{K}^{2}}}+990\,{\frac {u_{{x,x,x,x,x,x}}u_{{x,x,x}}}{K}}+132\,{\frac {
uu_{{x,x,x,x,x,x,x,x,x}}}{K}}
+132\,{\frac {u_{{x}}u_{{x,x,x,x,x,x,x,x}
}}{K}}+1995840\,{\frac {{u}^{5}u_{{x}}}{{K}^{5}}}+11880\,{\frac {uu_{{
x}}u_{{x,x,x,x,x,x}}}{{K}^{2}}}+990\,{\frac {u_{{x,x}}u_{{x,x,x,x,x,x,
x}}}{K}}+3326400\,{\frac {{u}^{3}u_{{x}}u_{{x,x}}}{{K}^{4}}}+55440\,{
\frac {u_{{x}}u_{{x,x}}u_{{x,x,x,x}}}{{K}^{2}}}
69300\,{\frac {u_{{x,x
,x}}{u_{{x,x}}}^{2}}{{K}^{2}}}+831600\,{\frac {{u}^{4}u_{{x,x,x}}}{{K}
^{4}}}+831600\,{\frac {u{u_{{x,x}}}^{2}u_{{x}}}{{K}^{3}}}+831600\,{
\frac {{u}^{2}u_{{x,x,x}}u_{{x,x}}}{{K}^{3}}}
+55440\,{\frac {uu_{{x,x}
}u_{{x,x,x,x,x}}}{{K}^{2}}}+332640\,{\frac {{u}^{2}u_{{x,x,x,x}}u_{{x}
}}{{K}^{3}}}+u_{{t}}+u_{{x,x,x,x,x,x,x,x,x,x,x}}+5940\,{\frac {{u}^{2}
u_{{x,x,x,x,x,x,x}}}{{K}^{2}}}
\end{math}

\section{Modified equations $E^n_p$}

We obtained sufficient conditions for two-soliton
solution when singularity order $p=2$. If $p$ changes, we still have
the same conditions ($\tau$-function conserves, but solution form changes).
So at this approach classification (tab. \ref{class}) is $p$-independent. For
example, we built modified 5th order SK equation with kink-type solitons \cite{Conte}
($p=1$)
\begin{equation}\label{mSK}
(\mathrm{SK})^5_1(u)=u_{{t}}+u_{{x,x,x,x,x}}+60\,{\frac {{u_{{x}}}^{3}}{{K}^{2}}}+30\,{
\frac {u_{{x,x,x}}u_{{x}}}{K}}
\end{equation}
From (\ref{mSK}) SK equation can be derived by taking $x$-derivative and substituting $u_x=v$. The same trick can be made for $p>2$, and that will lead us to differintegral equations. For example, modified 3th order Lax equation ($p=4$).
\begin{equation}
(\mathrm{Lax})^3_4(u)=
u_{{x,x,x}}+36\,{\frac {\int _{x}^{\infty }\!u \left( \xi,t \right) {d
\xi}u}{K}}+12\,{\frac {\int _{x}^{\infty }\!\int _{\eta}^{\infty }\!u
 \left( \xi,t \right) {d\xi}\,{d\eta}u_{{x}}}{K}}+u_{{t}}
\end{equation}
Solitary waves solutions for modified Lax and SK equations are represented at fig. below.
\begin{figure}[h!]\label{waves}
\centering
\includegraphics[scale=0.25]{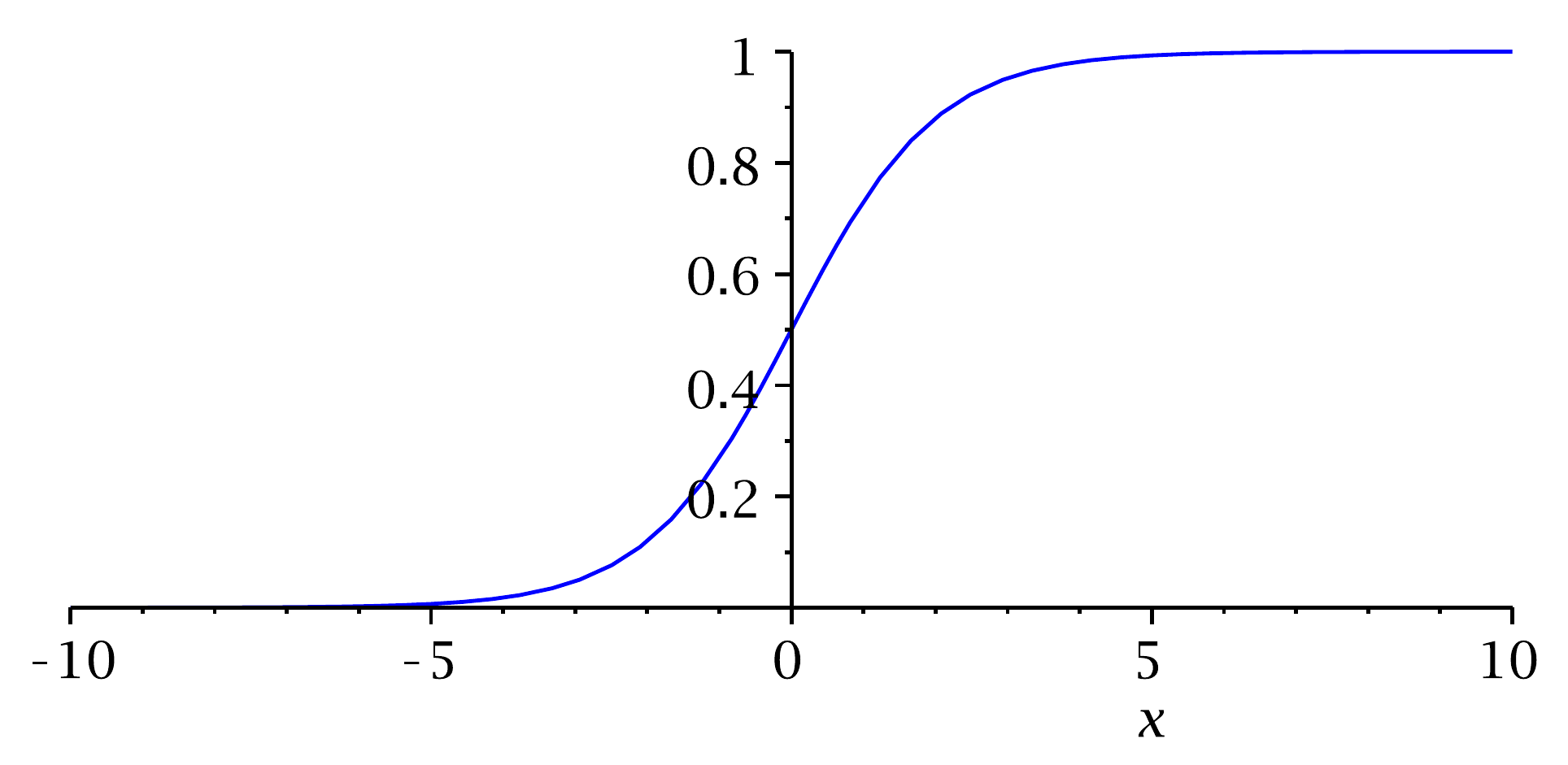}
\includegraphics[scale=0.25]{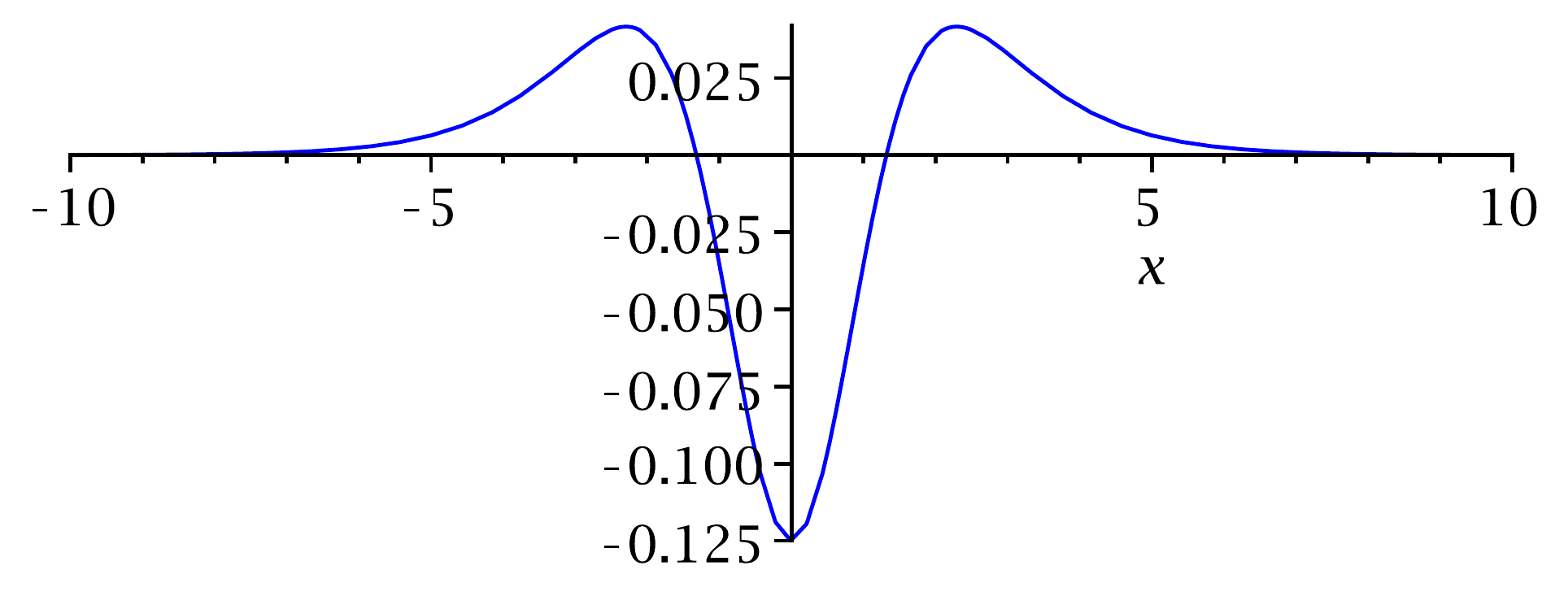}
\caption{Solitary waves for $(\mathrm{SK})^5_1(u)$ and $(\mathrm{Lax})^3_4(u)$ equation}
\end{figure}

\section{Conclusion}

Using the $\tau$-function substitution we classified high order PDEs with polynomial nonlinearities that accept two-soliton
solution in Hirota's form (tab. \ref{class}). All $E_p^n$ equations admit dispersion law in form $q_i=p_i^n$.
In integrable case ratio $\alpha_{12}$ uniquely determines single equation or hierarchy
of equations.
In paper \cite{Hietarinta} were established similar conditions for two-soliton solution for bilinear type equations.
But there was no information on general structure of these equations.
We discovered, apart from known completely integrable cases (Lax and SK hierarchies, \cite{Nuseir, Conte, Goriely, Saravi, Wazwaz_1, Wazwaz_2, Shen, Salas}), one hierarchy of two-soliton equations (new)
and countable family of isolated two-soliton equations (tail). 

Furthermore,
this algebraic structure does not depend on singularity order of an equation. So the transformation
of singularities that preserves form of a soliton should not produce another partial or complete integrable PDEs.

\end{document}